\documentclass[prb,twocolumn,groupedaddress,amsmath,floatfix]{revtex4}
\usepackage{graphicx}
\usepackage{bm}

\newcommand{\ec}{\varepsilon_{\rm c}}
\newcommand{\Ec}{E_{\rm c}}
\newcommand{\vf}{v_{\rm F}}
\newcommand{\vd}{v_{\Delta}}

\newcommand{\tc}{{T_{\rm c}}}

\newcommand{\phdag}{{\phantom{\dagger}}}
\newcommand{\ps}{{\phantom{*}}}

\newcommand{\kb}{k_{\rm B}}

\newcommand{\htunn}{H_{\rm T}}
\newcommand{\bk}{{\bf k}}
\newcommand{\bp}{{\bf p}}
\input{epsf}

\begin{document}
\title{Nodal Protectorate: A Unified Theory of the $ab$-plane and $c$-axis 
Penetration Depths of Underdoped cuprates}
\author{Daniel E.~Sheehy}
\altaffiliation[Present address: ]{Department of Physics, University of Colorado, Boulder, CO 80309, USA}
\author{T.P.~Davis}
\author{M.~Franz}
\affiliation{
Department of Physics and Astronomy, 
University of British Columbia, 
6224 Agricultural Road, Vancouver, B.C.~V6T1Z1, Canada}
\date{\today}
\begin{abstract}
We formulate a model describing the doping ($x$) and temperature ($T$)
dependence of
the $ab$-plane and $c$-axis penetration depth of a cuprate superconductor.
The model incorporates the suppression of the superfluid density 
with underdoping as the system approaches the Mott-Hubbard insulating state
by augmenting a $d$-wave BCS model with a phenomenological 
charge renormalization factor that is vanishingly small for states away 
from the nodes of the $d$-wave pair potential but close to unity in the 
vicinity of the nodes. The $c$-axis penetration depth is captured 
within a model of incoherent electron tunneling between the CuO$_2$ planes.  
Application of this model to the recent experimental data on the 
high-purity single crystals of YBa$_2$Cu$_3$O$_{6+\delta}$ implies 
existence of a ``nodal protectorate'', a $k$-space 
region in the vicinity of the nodes whose size decreases
in proportion to $x$, in which $d$-wave quasiparticles remain sharp even 
as the system teeters on the brink of becoming an insulator. The 
superfluid density, which is extremely small for these samples, also 
appears to come exclusively from these protected nodal regions.  
\end{abstract}
\maketitle

\section{Introduction}
\label{SEC:intro}
It is believed that to understand the high-$\tc$ cuprate superconductors, 
one must understand the problem of doping a Mott insulator.~\cite{anderson1}
Conversely, 
understanding the manner in which superconducting order gives way to the
antiferromagnetic insulator at half filling would provide important clues 
to the solution of this fundamental challenge.\cite{emery1,so5,balents1,ft1}
Indeed, those cuprates which have the lowest concentrations $x$ of dopants 
(so-called \lq\lq underdoped\rq\rq cuprates) 
possess many of the most enigmatic and mysterious experimental properties.  
In particular, as $x$ is reduced, the zero-temperature pair-potential 
maximum $\Delta_0 $ {\it grows\/} while the 
zero-temperature superfluid stiffness $\rho_s(0)$
and transition temperature $\tc$  {\it decrease\/}.\cite{REF:Uemura}  
In addition, the underdoped cuprates possess a ``pseudogap''\cite{timusk} 
in the single-particle excitation spectrum that persists above $\tc$.  
Taken together, these suggest that the way
superconductivity is destroyed as $x \to 0$ is very unusual.
Unfortunately, there has been a paucity of data on such very underdoped 
cuprates, in part due to sample preparation difficulties.  In addition, 
such materials are often very disordered, complicating analysis of 
their intrinsic physical properties.  Recently, however, Liang 
{\em et al.}~\cite{REF:Liang} have devised a way to vary $x$ in the doping phase 
diagram for a single crystal of YBa$_2$Cu$_3$O$_{6+\delta}$ (YBCO) in an
essentially continuous fashion.
 Briefly, their method utilizes the fact that the {\it effective} doping on 
the CuO$_2$ planes is governed by both
the oxygen concentration and the degree of oxygen ordering in the CuO chains.
By varying the latter variable (via room temperature annealing), 
these authors can alter the effective doping (and thus $\tc$) to explore 
the doping dependence of physical quantities (such as 
the $c$-axis penetration depth~\cite{REF:Hosseini,prl}) in the strongly underdoped
regime $x\to 0$. 

Motivated by the recent c-axis penetration  results of Hosseini 
{\em et al.},\cite{REF:Hosseini,prl} in  this Paper 
we develop a model aimed at capturing the overall phenomenology of the doping
and temperature dependence of the $ab$-plane and $c$-axis 
penetration depths ($\lambda_{ab}$ and $\lambda_c$, respectively). 
Our model combines incoherent tunneling in the 
$c$-direction with a phenomenological momentum-dependent charge 
renormalization factor inspired by the 
idea of Ioffe and Millis~\cite{REF:Ioffe} to account 
for anomalous doping dependences. On a more fundamental level we are 
interested 
in understanding the constraints that this data imposes on the models 
of strongly correlated electron matter describing the $d$-wave 
superconductor on the verge of becoming a Mott-Hubbard insulator.
Our results indicate that the appropriate effective theory must exhibit a 
``nodal protectorate'' consisting of regions in the vicinity of the 
$d$-wave gap nodes where the quasiparticles remain well defined even as 
$x\to 0$.   

To begin, let us review the overall phenomenology of the penetration depth 
in the cuprates.  For the moment, 
we are interested in general trends (especially in the strongly underdoped 
regime $x \to 0$) in the penetration depth.
To the extent that such data is available,
\cite{REF:Uemura,REF:Bonn96,REF:Lemberger}
the $ab$-plane penetration depth $\lambda_{ab}(x,T)$
exhibits the following behavior:
\begin{equation}
\rho_s^{ab}(x,T)=\frac{\hbar^2 c^2 d}{16\pi e^2 \lambda^{2}_{ab}(x,T)}
\simeq a x-bk_BT,
\label{eq:slopeab}
\end{equation}
where we have expressed $\lambda_{ab}(x,T)$ in terms of the associated 
superfluid stiffness $\rho_s^{ab}(x,T)$ measuring
the free energy cost to a variation of the order-parameter phase in the 
$ab$ plane. Here, $d$ denotes the distance between copper-oxygen layers. 
For YBCO the relevant parameters are $a\simeq 244$meV and $b\simeq
3.0$.\cite{REF:Bonn96,REF:Lee}
We shall assume that this phenomenology holds for all underdoped cuprates
\cite{tami1,remark0}.  
The coefficient of $T$ in Eq.~(\ref{eq:slopeab}) is {\it nearly} doping 
independent, in the sense that although
it varies as the gap amplitude $\Delta_0$ varies, it does not vary as 
strongly as $\rho_s^{ab}(x,0)$ i.e., linear in $x$. The $T$-linear term 
is well known and is understood to be due
to the excitation of quasiparticles out of the condensate near the nodes of 
the $d$-wave order parameter.\cite{REF:Scalapino} 
The fact that this relation holds to very low dopings 
(along with other experiments  sensitive to nodal physics, see, e.g., 
Ref.~[\onlinecite{REF:Sutherland}]) indicates that near the nodes elementary
excitations in cuprates are well described by conventional BCS quasiparticles
even for $x\to 0$.  

The doping dependence of $\rho_s^{ab}(x,T)$, on the other hand, represents
one of the central mysteries in the cuprates.\cite{REF:Lee} 
The linear in $x$ reduction of 
$\rho_s^{ab}(x,0)$ as one approaches half filling must be attributed to the 
proximity of the Mott-Hubbard insulating phase. However, most theoretical 
treatments (such as the RVB-type\cite{rvb} and Gutzwiller 
projection\cite{projected,laughlin} approaches) that can account for the 
observed $x$-dependence of $\rho_s^{ab}(x,0)$
also predict a strongly $x$-dependent prefactor $b$ in Eq.~(\ref{eq:slopeab}),
in disagreement with the data. Thus, the central theoretical 
problem appears to lie in constructing a model that would make only a small 
fraction $\sim x$ of all electrons
participate in the superconducting condensate while at the same time 
preserve the simple BCS character of the 
nodal quasiparticles. In the following
we resolve this problem by postulating that, at least for the purposes
of studying the superfluid stiffness, the underdoped cuprates can be 
described by the BCS theory augmented with a phenomenological constraint
that only Cooper pairs composed of electrons with momenta in the vicinity
of the nodal points effectively couple to the external electromagnetic 
field. We implement this idea by extending the ``effective charge 
renormalization'' concept introduced in this context by Ioffe and 
Millis.\cite{REF:Ioffe}
\begin{figure*}
\includegraphics[width = 12.8cm]{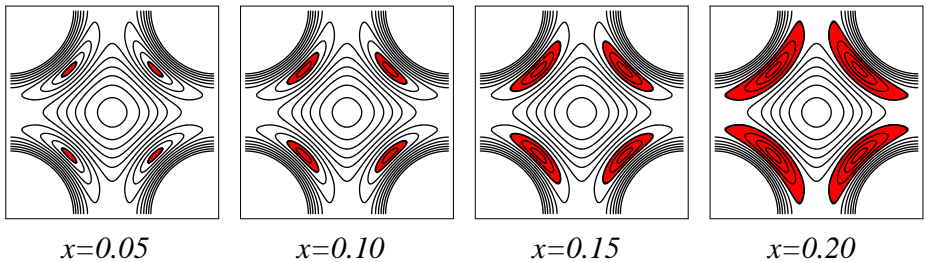}
\caption{Schematic plot of the assumed form for $Z_{\bf k}$, showing the 
``nodal protectorate'' region (shading) of the Brillouin zone where states 
contribute to the formation of the condensate for a cuprate at a particular 
doping $x$.  The black lines are the constant energy contours in 
the Brillouin zone (which do not vary with doping).  Near optimal 
doping ($x=20$), electrons 
in a large region around the node contribute to the Meissner response.  
As the doping is 
reduced this region is progressively reduced, leaving a small ``patch'' 
near the nodes where superconductivity remains robust.  We remark that 
the c-axis penetration depth measurements of Ref.~\onlinecite{prl} were 
performed on extremely underdoped samples with effective 
dopings $x$ that are approximately represented by the leftmost panel.
}
\label{fig6}
\end{figure*}

This model, described in a more detail below, is {\em designed} to reproduce 
the $ab$-plane 
phenomenology embodied in Eq.\ (\ref{eq:slopeab}). What makes us believe
that it might be of more general validity is the fact that it also 
reproduces the doping dependence of the $c$-axis superfluid stiffness,
$\rho_s^c(x,T)$, once we adopt a model for interlayer coupling that gives
the correct (nonlinear) temperature dependence of this quantity. 
  The phenomenology along the $c$-axis is tantalizingly similar to the $ab$ 
plane case and can be written in the following way:
\cite{REF:Hosseini,prl,REF:Hosseini98,panag1}
\begin{equation}
\rho_s^{c}(x,T)\propto\lambda^{-2}_{c}(x,T) \simeq A x^\alpha-BT^\alpha,
\label{eq:slopec}
\end{equation}
where the exponent $\alpha$ is, within the experimental accuracy, the same 
for both $x$ and $T$ and close to $2$, while $A,B$ are again $x$ and $T$
independent constants. The fact that the temperature 
dependence is nearly quadratic (as opposed to linear)
points to the {\em incoherent} coupling between the copper-oxygen planes,
as discussed by previous authors.\cite{REF:Radtke0,REF:Radtke} In the 
following we shall clarify the specific conditions under which such a nearly quadratic $T$-dependence arises for the incoherent tunneling model. More importantly we show that within our scheme for the effective 
charge the {\em same power law} also necessarily governs the doping 
dependence of $\rho_s^{c}(x,0)$, in agreement with Eq.\ (\ref{eq:slopec}). 

This Paper is organized as follows: In Sec.~\ref{SEC:ab}, we augment the standard BCS expression for 
$\lambda_{ab}$ with a phenomenological charge renormalization factor.  By appealing to 
experimental data, the parameters determining this factor are extracted.  In Sec.~\ref{SEC:caxis}, we consider
a tunneling model of $c$-axis transport and augment the resulting expression for $\lambda_c$ 
 with the same charge renormalization factor in an analagous way.  It is shown that the $T$ and $x$ 
dependences of our expression for $\lambda_c$ agree qualitatively (having made certain assumptions) with
Eq.~(\ref{eq:slopec}).  In Sec.~\ref{SEC:caxis}, we fit our expression for $\lambda_c$ to the results 
of Refs.~\onlinecite{REF:Hosseini,prl}).  In Sec.~\ref{SEC:concl} we make some concluding remarks.  Certain
calculational details are relegated to Appendices ~\ref{a1} amd ~\ref{a2}.


\section{$ab$-plane properties}
\label{SEC:ab}

The starting point of our calculation of  the $ab$-plane penetration depth 
$\lambda_{ab}(T)$ is the following Hamiltonian for
a $d$-wave superconductor coupled to an applied in-plane electromagnetic field:
\begin{eqnarray}
\label{eq:inplaneham}
H&=& H_{\rm pair} + H_{\rm int},
\\
H_{\rm pair} &=& \sum_{{\bf k},\sigma} \epsilon_{\bf k} 
c_{{\bf k},\sigma}^{\dagger} c_{{\bf k},\sigma}^{\phdag}+\sum_k 
\Delta_{\bf k} (c_{{\bf k},\uparrow}^{\dagger} c_{-{\bf k},\downarrow}^{\dagger} + 
{\rm h.c.} ),\nonumber
\end{eqnarray}
where $H_{\rm pair}$ is the standard BCS pairing Hamiltonian with
the single-particle energy $\epsilon_{\bf k} = 
-2t(\cos k_x a +\cos k_y a ) -\mu$ 
with $a$ the lattice spacing. The pair potential $\Delta_{\bf k}
=\frac{1}{2}\Delta_0(\cos k_x a -\cos k_y a )$; we shall always assume
that the maximum pair potential $\Delta_0$ is approximately temperature 
independent. 
Here, $H_{\rm int}$ refers to additional interactions not captured by 
$H_{\rm pair}$ which describe the physics of the proximity to the Mott-Hubbard
insulating phase at half filling. Our strategy will be to compute 
$\lambda^{-2}_{ab}$  by first neglecting
$H_{\rm int}$ and computing the usual BCS result for the superfluid density
in a $d$-wave superconductor.  Then,
we shall include the residual interactions  $H_{\rm int}$ via a 
phenomenological charge renormalization motivated by the work of 
Millis {\em et al.}~\cite{REF:Millis98} and Ioffe and 
Millis.~\cite{REF:Ioffe} 
 The calculation of the penetration depth for $H_{\rm pair}$ 
is standard and is presented in Appendix~\ref{a1} for completeness.  
The final result may be expressed as
\begin{eqnarray}
\frac{1}{\lambda^2_{ab}(x,T)} &=& 
\frac{e^2\,n}{d}
\sum_{\bf k}  Z_\bk^2\left( 
\frac{\partial\epsilon_{\bf k}}{\partial k_x}\right)^2 
\frac{\Delta_{\bf k}^2}{E_{\bf k}^2} 
\nonumber \\
&&\quad \times
\left[ \frac{1}{E_{\bf k}} - \frac{\partial}{\partial E_{\bf k}} \right]
\tanh\frac{1}{2}\beta E_\bk,
\label{eq:standardab}
\end{eqnarray}
with $E_\bk=\sqrt{\epsilon_\bk^2+\Delta_\bk^2}$ the BCS $d$-wave excitation
spectrum  and $\beta=1/T$ the inverse temperature.  The factor $Z_{\bf k}^2$ 
will be discussed shortly.

We shall take Eq.~(\ref{eq:standardab}) to be the starting point of our 
phenomenological analysis of the $ab$-plane penetration depth data. It 
contains a sum over all momentum vectors ${\bf k}$ in the Brillouin zone
and has been cast into a form where the zero 
temperature value $\lambda_{ab}(0)^{-2}$ and the finite temperature correction
$\delta\lambda_{ab}(T)\equiv \lambda_{ab}^{-2}(0)-\lambda^{-2}_{ab}(T)$ are 
treated on an equal footing; in particular they 
are both dominated by the regions in the $k$-space close to the nodes of
the $d$-wave gap function.  
As discussed in Appendix~\ref{a1}, for $Z_{\bf k} = 1$ this expression is 
fully equivalent to well-known expressions
 for $\lambda^2_{ab}$ that have appeared previously 
in the literature.~\cite{scalapino1,dhlee1} 

Following Ref.~[\onlinecite{REF:Ioffe}] we have introduced in 
Eq.~(\ref{eq:standardab}) a phenomenological 
charge renormalization factor $Z_{\bf k}$ by means of the replacement  
\begin{equation}
e^2\sum_\bk\to e^2\sum_\bk Z_{\bf k}^2.
\end{equation}
This is done to incorporate 
the effects of the interactions contained in $H_{\rm int}$ 
responsible for the gradual demise of superconducting order as  
the Mott-Hubbard insulating phase is approached near the  half filling. 
The main difference between our approach and that of  
Ref.~[\onlinecite{REF:Ioffe}] is that we have incorporated $Z_{\bf k}$ into 
the full
expression for $\lambda^2_{ab}$, as opposed to the temperature dependent 
part only. This is in keeping with our philosophy of treating both 
components on the equal footing. At this stage we do not attempt to justify
the inclusion of $Z_\bk$ from microscopic considerations; we merely state 
that such charge renormalization is not prohibited by any general principle
(and is known to occur e.g.\ in the Fermi liquid
theory). We offer some discussion of the possible origins of $Z_\bk$
in Sec.\ IV.

In the absence of a microscopic theory for $Z_\bk$ we must rely
on experimental data to infer the behavior of 
this quantity. We begin by reiterating that, in
 weak coupling  BCS theory $Z_\bk=1$ and 
a direct evaluation of Eq.~(\ref{eq:standardab}) yields a result
which does not conform to the observed $ab$-plane 
phenomenology~\cite{REF:Lee}. 

In particular BCS theory yields a correct
linear-$T$ dependence of $\delta\lambda_{ab}(T)$ but the wrong doping 
dependence of 
$\lambda_{ab}^{-2}(0)$, which would scale with the total number of electrons 
$(1-x)$ in disagreement with Eq.~(\ref{eq:slopeab}).
To rectify this discrepancy the form of $Z_\bk$ must be modified.
Since at low temperatures the $T$-dependence of $\delta\lambda_{ab}(T)$ 
comes from thermally
excited quasiparticles near the four nodal points of a $d$-wave gap,
it is clear that in order to preserve this correct temperature dependence
$Z_\bk$ must remain {\em close to unity} in the nodal 
regions. To suppress the overall magnitude of  $\lambda_{ab}^{-2}(0)$ it then 
follows that $Z_\bk$ must be small {\em outside the nodal regions}. In 
addition, to conform to the $\lambda_{ab}^{-2}(0)\sim x$ scaling the
size of the ``patch'' in which $Z_\bk\approx 1$ must scale with $x$.

The $ab$-plane phenomenology thus dictates the following form for the charge 
renormalization factor
\begin{equation}
Z_\bk\approx\biggl\{
\begin{array}{ll}
Z_0 & {\rm for} \ E_\bk<\Ec, \\
0 & {\rm for} \ E_\bk>\Ec,
\end{array} 
\label{eq:z1}
\end{equation}
with $Z_0$ a constant of order 1 and $\Ec\sim x$ the characteristic 
cutoff energy that is chosen to obtain the correct magnitude of the 
zero-temperature superfluid stiffness. As illustrated in Fig.\ \ref{fig6},
this choice of $Z_\bk$ effectively restricts the sum over the entire 
Brillouin zone in Eq.~(\ref{eq:standardab}) to the immediate vicinity of 
the nodes. At low 
temperatures, such a choice affects the diamagnetic response (which is 
sensitive to the entire Fermi surface)
but not the paramagnetic response which governs the temperature dependence.

To see how this reproduces the assumed phenomenology,
we proceed by inserting the assumed form for $Z_\bk$ 
into Eq.~(\ref{eq:standardab}). We make 
an assumption, which we verify momentarily, that $\Ec$ is sufficiently 
small (i.e.\ $\Ec\ll \Delta_0$) 
for underdoped cuprates that it is permissible to linearize 
$\epsilon_\bk$ and $\Delta_\bk$
in the vicinity of the four nodes.\cite{REF:Durst} We implement this
linearization  by introducing
a local coordinate system $(k_1,k_2)$ centered at the nodal point
with axes rotated $45^\circ$ with respect to $(k_x,k_y)$.  Expanding 
to leading order, we have  
$\epsilon_\bk\to v_Fk_1$, $\Delta_\bk\to v_\Delta k_2$ with $v_F$, 
$v_\Delta$ being the Fermi and gap velocities at the node, respectively. 
We thus obtain
\begin{eqnarray}
\frac{1}{\lambda^2_{ab}} 
&=& \frac{2 e^2Z_0^2v_F^2 n}{d} \int_{E_\bk<\Ec}
\frac{dk_1 dk_2}{(2\pi)^2} \frac{\vd^2k_2^2}{E_\bk^2} \nonumber \\
&\times&
\left[\frac{1}{E_\bk}\tanh\frac{1}{2}\beta E_\bk
-\frac{1}{2}\beta{\rm sech}^2\frac{1}{2}\beta E_\bk
\right],
\end{eqnarray}
where $E_\bk=\sqrt{v_F^2k_1^2+\vd^2 k_2^2}$ is the linearized ``Dirac'' 
spectrum  of quasiparticles.
It is now useful to rescale the integration variables
as $v_Fk_1\to k_1$ and $\vd k_2\to k_2$ and pass to polar coordinates.
The angular integral is trivial and we obtain (inserting $\hbar$ and
$k_B$) 
\begin{subequations}
\begin{eqnarray}
\label{eq:abfin}
\!\!\!\frac{1}{\lambda^2_{ab}} 
&=& \!\frac{2 e^2Z_0^2n\vf}{\hbar^2 d\,\vd} 
\!\!\int_0^{\Ec} \!\frac{ dk}{4\pi}\!\!
\left[
\tanh\!\frac{\beta k}{2}\! -\! \frac{\beta k}{2}{\rm sech}^2 \frac{\beta k}{2} \!
\right],  \\
&\simeq& 
\frac{e^2Z_0^2n}{ 2\pi \hbar^2d }  \frac{\vf}{\vd}
\left( \Ec - 4 k_BT \ln 2 \right), \label{eq:analytic}
\end{eqnarray}
\end{subequations}
where the second line  applies in the regime $T\ll \Ec$.  

\begin{figure}
\includegraphics[width = 7.8cm]{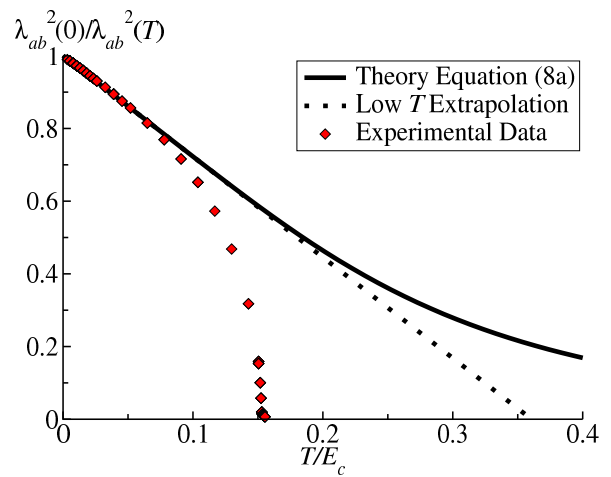}
\caption{Plot of normalized superfluid stiffness 
$(\lambda_{ab}(0)/\lambda_{ab}(T))^2$ with $\lambda_{ab}(T)$ given by 
Eq.~(\ref{eq:abfin}) (solid line) and Eq.~(\ref{eq:analytic}) 
(dashed line), respectively, as a 
function of the normalized temperature $T/E_{\rm c}$. Although they 
agree well at low temperatures, the deviation of 
Eq.~(\ref{eq:analytic}) from Eq.~(\ref{eq:abfin})  becomes significant 
at high temperatures.  As noted in the text, Eq.~(\ref{eq:abfin})
never reaches zero, indicating that the quasiparticle-excitation mechanism 
of depleting the condensate can never fully 
destroy superconductivity if the pair potential remains nonzero. Diamonds
represent the $ab$-plane data (YBCO 6.6) of Ref.\ [\onlinecite{REF:Bonn96}] 
scaled in such a way that the low-$T$ linear part agrees with the
theoretical model. 
}
\label{fig1}
\end{figure}

It is clear that if we take $\Ec \propto x$, then this reproduces Eq.~(\ref{eq:slopeab}).  
To make a rough estimate of how $\Ec$ must vary
with the $T_c$ of an underdoped sample for this picture to apply, 
we simply assume that $\tc$ may be determined by the $T$ at which 
$\lambda^{-2}_{ab}(T)$ in Eq.~(\ref{eq:analytic}) equals $ 0$.  
This criterion, which was also used by Lee and Wen,\cite{REF:Lee} gives
$\Ec \simeq 2.8 \kb\tc = 0.24 \tc$meV/K. 
Comparison with the experimental data shown in Fig.~\ref{fig1}
 illustrates that this overestimates $T_c$ by about a factor 
of 2.3 due to the fact that the actual data deviates significantly from 
the straight 
line at higher temperatures. To account for this deviation, and for future 
reference, we revise our estimate to  read
\begin{equation}
\Ec \simeq 6.4 \kb\tc = 0.55 \tc {\rm meV/K}.
\label{Ec1}
\end{equation}
This implies that even at optimal doping, $T_c\simeq 90$K, the energy 
scale $\Ec$ is of the order of maximum gap $\Delta_0\simeq 45$meV. This 
validates our assumption that for underdoped YBCO the linearized 
approximation should lead to quantitatively correct results.
Similarly one can estimate the doping dependence of $\Ec$ by comparing 
Eqs.\ (\ref{eq:slopeab}) and (\ref{eq:analytic}). For YBCO this yields
\begin{equation}
\Ec \simeq 2\pi a\frac{v_\Delta}{v_F}Z_0^{-2}x\approx [219.0{\rm meV}]x,
\label{Ec2}
\end{equation}
where we have taken $Z_0^2(v_F/v_\Delta)=7$, a value relevant to this 
material.\cite{REF:Ioffe}
In Sec.~\ref{SEC:fits} we shall see that a similar value of 
$\Ec$ also provides the best fit to $c$-axis penetration
depth data.  
 
There is an important caveat regarding the preceding analysis due to the fact that it relies on 
on the approximate formula Eq.~(\ref{eq:analytic}).  Indeed, 
the exact formula Eq.~(\ref{eq:abfin}) does not possess a $\tc$ in the sense of 
Eq.~(\ref{eq:analytic}): One can easily 
see that $\lambda_{ab}^{-2}(T) >0$  for any $T$. This is because we have 
taken $\Delta_{\bf k}$ to be nonzero and temperature-independent as 
implied for underdoped cuprates
by various experiments such as ARPES\cite{campuzano1} or tunneling.
\cite{renner1} For non-zero pair potential within  mean-field 
theory, the condensate is depleted only through thermal excitation of
quasiparticles; this leaves a small fraction of electrons in the condensate
at arbitrary temperatures. In other words within  mean-field 
BCS theory the only
way to kill the condensate is to drive    $\Delta_{\bf k}$ to zero.
In Fig.~(\ref{fig1}), we plot Eq.~(\ref{eq:abfin}) (solid line) and 
Eq.~(\ref{eq:analytic}) (dashed line) as a function of 
the normalized temperature $T/E_{\rm c}$. One can show that the solid line
approaches zero only asymptotically as $1/T^3$.  In any real system, once 
the superfluid stiffness becomes sufficiently small due to
quasiparticle excitations, fluctuation effects (e.g., phase 
fluctuations~\cite{emery1}) will rapidly  destroy the condensate.  
Thus, it is 
reasonable to assume that the relationship ({\ref{Ec1}) holds approximately 
for real systems; i.e., it provides an 
estimate for how the physical $\tc$ depends on the zero-temperature 
superfluid stiffness (parameterized by $\Ec$).
In the next section, we compute the $c$-axis superfluid stiffness within a 
tunneling model and will find similar behavior.

Finally we may ask whether there is another form of $Z_\bk$ that might
agree with the experimental data. One possibility is that, for some reason,
$Z_\bk$ does not vanish outside the nodal patch but goes to a small value
proportional to $\sqrt{x}$. This would clearly reproduce the 
observed $ab$-plane doping dependence.\cite{remark1}
 We shall see below, however, that such a 
form would {\em not} produce the correct doping dependence for $\lambda_c$;
the latter appears to dictate that $Z_\bk$ vanishes outside the patch.
Also, we have only considered situations when $Z_\bk$ depends on the momentum
through the energy $E_\bk$. While this assumption appears very natural one
can envision scenarios in which $Z_\bk$ would depend on the individual 
components of $\bk$.

\section{$c$-axis properties}
\label{SEC:caxis}

In the present 
section, we apply the philosophy of Sec.~\ref{SEC:ab} to the problem of the 
$c$-axis penetration depth, for which new data in the strongly underdoped 
regime has recently been obtained.\cite{REF:Hosseini,prl}  

\subsection{Tunneling Hamiltonian}
To model the electronic transport in the $c$-direction, we adopt the 
following tunneling Hamiltonian:
\begin{equation}
\htunn = \sum_{m,\sigma} \int d^2 r 
( t_{\bf r} c_{{\bf r},m+1,\sigma}^{\dagger} c_{{\bf r},m,\sigma}^{\phdag}
+ {\rm h.c.}),
\label{eq:htunn}
\end{equation}
where $c_{{\bf r},m,\sigma}$ annihilates an electron at position ${\bf r}$
in layer $m$ with spin projection $\sigma$. 
The matrix element $t_{\bf r}$ connects states in adjacent layers and is 
distinct from the quantity $t$ in Sec.~\ref{SEC:ab}.
The total system Hamiltonian, then, consists of a sum of intralayer Hamiltonians 
given by Eq.~(\ref{eq:inplaneham}) coupled by $\htunn$.  

To compute the $c$-axis superfluid
stiffness, we derive via the Kubo formula the response to an AC
electric field $E({\bf r},m,t)= E({\bf r},m){\rm e}^{i\omega t}$ 
in the $c$-direction.  To do this, we recall that such an electric field
may be incorporated via the Peierls substitution
\begin{equation}
t_{\bf r} \to t_{\bf r}{\rm e}^{i\frac{ed}{c} A({\bf r},m)},
\end{equation}
where $A({\bf r},m) = \frac{ic}{\omega} E({\bf r},m){\rm e}^{i\omega t}$
and $d$ is the interlayer spacing.
The $c$-axis current density $j({\bf r},m)$ is given by
\begin{eqnarray}
&&j({\bf r},m) = -ie t_{\rm r}  
(c_{{\rm r},m+1,\sigma}^{\dagger} c_{{\rm r},m,\sigma}^{\phdag}-
 {\rm h.c.})
\nonumber 
\\
&&\qquad + \frac{e^2 d}{c} t_r A({\bf r},m) 
(c_{{\rm r},m+1,\sigma}^{\dagger} c_{{\rm r},m,\sigma}^{\phdag}+
 {\rm h.c.}).
\label{eq:j}
\end{eqnarray}
The two terms in Eq.~(\ref{eq:j}) have the familiar form of the 
paramagnetic and diamagnetic contributions to the current, respectively.
We proceed to compute the associated contributions
to the conductivity to leading order in perturbation theory in
the matrix element $t_{\bf r}$. We shall furthermore assume that the in-plane Green functions
are given by the usual $d$-wave BCS forms:
%
%
\begin{subequations}\begin{eqnarray}
\langle T_{\tau} c_{-{\bf p},m,\downarrow}^{\phdag}(\tau)
 c_{{\bf p},m',\uparrow}^{\phdag}(0)
\rangle &=& \delta_{m,m'} F({\bf p},\tau) ,
\\
\langle T_{\tau} c_{{\bf p},m,\sigma}^{\phdag}(\tau)
 c_{{\bf p},m',\sigma}^{\dagger}(0)
\rangle &=& -\delta_{m,m'}  G({\bf p},\tau),
\end{eqnarray}\end{subequations}
with $G({\bf p},\omega)$ and $F({\bf p},\omega)$ given in Eq.~(\ref{eq:g}) and Eq.~(\ref{eq:f}), respectively.

Of principal importance is our choice for the form of the tunneling matrix element  
$t_{\bf r}$.  Allowing for nontrivial ${\bf r}$-dependence is essential, as 
a spatially uniform $t_{\bf r}$ (so that in-plane momentum is conserved 
during the tunneling process) yields a $T$-linear correction to the penetration depth at low 
temperatures.  This may be understood by noting that such purely coherent 
tunneling between layers simply yields a three-dimensional $d$-wave 
superconductor.  As noted previously,\cite{REF:Radtke0,REF:Radtke}
 the absence of such linear 
behavior points towards incoherence in the $c$-axis tunneling.
Before proceeding, we remark that another possibility to obtain non-linear-$T$ 
dependence is the proposal of Xiang and Wheatley.\cite{REF:Xiang}  They
find that by including a momentum dependence to the 
tunneling matrix element (arising from the structure of the copper and
oxygen orbitals
involved in tunneling), one finds a $T^5$ dependence in disagreement with the 
overall observed temperature dependence [Eq.~(\ref{eq:slopec})].
It is possible that such a $T$-dependent contribution is present but simply
masked by the dominant $T^2$ behavior; however, we shall not include this
possibility in the following.  

Our strategy for finding a form for $t_{{\bf p}-{\bf k}}$ that  gives the 
experimentally observed $T$-dependence is to assume that interplane disorder 
scatters the momentum states tunneling from layer to layer.  We assume that the 
disorder-averaged matrix element   $\overline{t_\bk} = 0$.  
 For the disorder-averaged product of tunneling matrix elements we choose
\begin{subequations}\begin{eqnarray}
\overline{t^{*}_{\bf k}t^{\ps}_{{\bf k}+{\bf q}}}
&=& (2\pi)^2 \delta({\bf q}) {\cal T}^{2}_{k},
\\
\label{eq:t2}
{\cal T}^2_{\bf k} &=& 
\frac{t_\perp^2}{\pi \Lambda^2}{\rm e}^{-k^2/\Lambda^2},
\end{eqnarray}\end{subequations}
with $t_\perp$ being an energy scale characterizing the strength of tunneling
and $\Lambda$ being an inverse length scale characterizing 
the degree of momentum non-conservation. We expect
that the specific form of ${\cal T}^2_{\bf k}$ is unimportant as long
as it includes the possibility of tunneling between different in-plane 
momentum states.
%
Similar models that incorporate disorder in c-axis transport have
been considered by the authors of 
Refs.~[\onlinecite{REF:Radtke0,REF:Radtke}] who also find $T^2$ behavior.  
In Ref.~[\onlinecite{REF:Radtke}], it is assumed that
${\cal T}_\bk$ depends only on the component of $\bk$ {\em parallel}
to the Fermi surface (implying complete non-conservation of the 
{\em perpendicular} component).
Under these assumptions, the leading 
temperature dependence of $\lambda_c^{-2}(T)$ is quadratic, provided 
that the parallel component of $\bk$ is conserved. It is however
not easy to envision an interlayer scattering mechanism that would produce
tunneling that is perfectly conserving for the momentum component
parallel to the Fermi surface while totally non-conserving in the 
perpendicular direction. 
A much more natural assumption, embodied by 
our choice Eq.~(\ref{eq:t2}), is to take
${\cal T}_\bk$ to be {\em isotropic} in the $ab$-plane. 
In the next section, we turn to the evaluation of the c-axis penetration depth
with this choice of tunneling matrix element.  
We will show that
even  isotropic tunneling produces a nearly $T^2$ dependence of  
$\delta\lambda_c(T)$ as a crossover behavior when the anisotropy of the Dirac
spectrum near the node is taken into account. The same physics also 
produces the nearly $x^2$ doping dependence of $\lambda_c^{-2}(0)$, provided 
that we implement the charge renormalization factors as we did
in the Sec.~\ref{SEC:ab}



\subsection{$c$-axis penetration depth}
The calculation of the $c$-axis penetration depth within the tunneling
Hamiltonian 
formalism parallels that presented in Appendix A for $\lambda_{ab}$. The details can be 
found in Ref.~[\onlinecite{REF:Radtke}] and the result is given by 
\begin{equation}
\frac{1}{\lambda_c^2} =   8e^2 d  \sum_{\bf k,\bf p} 
 {\cal T}^2_{{\bf k}-{\bf p}}T\sum_{i\omega}
F({\bf k},\omega)  F({\bf p},\omega),
\label{eq:jos}
\end{equation}
 As noted in Ref.~[\onlinecite{REF:Radtke}], 
this is equivalent to the standard result~\cite{REF:Mahan} for the critical 
Josephson current through a weak link.  This 
is sensible, since the way in which supercurrent is passed in the 
$c$-direction for weakly coupled layers is via the Josephson effect.

We now augment Eq.~(\ref{eq:jos}) in the same manner as in 
Sec.~\ref{SEC:ab}, by  making the replacement 
\begin{equation}
e^2\sum_{{\bf k},{\bf p}}\to e^2\sum_{{\bf k},{\bf p}}Z_{\bf k} Z_{\bf p}.
\end{equation} 
With the choice made for $Z_{\bf k}$ in Eq.\ (\ref{eq:z1}) this 
again amounts to  restricting the summations over ${\bf k}$ and 
${\bf p}$ to within the vicinity of the nodes. 
Evaluating the sum over Matsubara frequencies, we obtain 
\begin{eqnarray}
\label{eq:lambdac}
\frac{1}{\lambda_c^2} &=& 2e^2 d \sum_{{\bf k},{\bf p}} 
{\cal T}^2_{{\bf k}-{\bf p}}Z_{\bf k} Z_{\bf p}
\frac{\Delta_{\bf k} \Delta_{\bf p}}{E_{\bf k} E_{\bf p}} 
\nonumber \\
&\times&
\biggl[  \frac{\tanh\frac{1}{2}\beta E_\bk + \tanh\frac{1}{2}\beta E_\bp}
{E_\bk+E_\bp} \nonumber \\
&&\quad -
\frac{\tanh\frac{1}{2}\beta E_\bk - \tanh\frac{1}{2}\beta E_\bp}{E_\bk-E_\bp}
\biggr].
\end{eqnarray}
This equation contains a four dimensional momentum integral and one needs
to employ numerical methods to obtain the full $T$ and $\Ec$ dependences
of $\lambda_c$. We shall present such numerical results shortly. 
In order to elucidate the physics contained in this expression we first
study the leading behavior analytically using scaling arguments in the
limit of low $T$ and low doping $x$ [entering via $\Ec \propto x$ as in 
Sec.~(\ref{SEC:ab})] when the physics is dominated by the
nodal regions.   

\begin{figure}
 \epsfxsize=\columnwidth
\centerline{\epsfbox{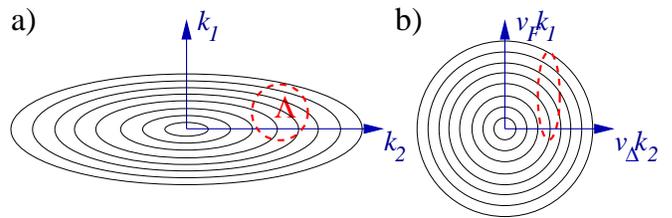}}
\vskip0.50cm \caption{Schematic plot of the constant energy contours in 
momentum space in the vicinity of a node. On the left, contours at energy $E_{\bf k}$ satisfy  
$E_{\bf k}^2 = \vf^2k_1^2 + \vd^2 k_2^2$ (i.e., they are 
ellipses).  The tunneling
matrix element Eq.~(\ref{eq:t2}) conserves momentum within a range 
$\Lambda$ represented by the dashed circle.  
By rescaling the plot so that the axes are $\vf k_1$ and $\vd k_2$, the 
constant energy contours are circles, but the 
circle representing the degree of momentum conservation has become 
distorted, indicating that the $k_1$ component
of the quasiparticle momentum is effectively conserved to a lesser 
degree than $k_2$.  
}
\label{fig7}
\end{figure}

\subsection{$c$-axis penetration depth: scaling analysis}
Consider the low-$T$ behavior of $\lambda_c^2$ 
for $T\ll \Ec\ll\Delta_0$. We shall denote the 
$T$-dependent correction to the stiffness by
$\delta\lambda_c(T) \equiv \lambda^{-2}_c(0) -  \lambda^{-2}_c(T)$.  
An explicit expression for $\delta\lambda_c(T)$ 
may be obtained from Eq.~(\ref{eq:lambdac}})  by replacing each $\tanh$ function
by $1-\tanh$ (with the same argument).  
In the low-$T$ limit, these functions restrict the momentum integrals to the  
vicinity of the nodes, so that we  
can take $Z_\bk\simeq Z_0$ everywhere. Linearizing near 
the four nodes and rescaling to the natural momentum variables as in 
Eq.~(\ref{eq:abfin}) we find that the tunneling matrix element 
${\cal T}^2_\bk$ becomes {\em anisotropic}, as illustrated in Fig.~\ref{fig7}.
In these natural ``nodal'' variables the momentum component perpendicular
to the Fermi surface ($\vf k_1$) is conserved to a lesser degree than the 
parallel component ($\vd k_2$). This gives rise to three distinct 
scaling regimes depending on the magnitude of $T$ with respect to the
two natural scales of the problem, $\vf\Lambda$ and $\vd\Lambda$. In what 
follows, we shall assume that $\vd \ll \vf$, as is known
to be the case for cuprates.~\cite{REF:Sutherland}

(i) For $\vd\Lambda \ll v_F\Lambda\ll T$ thermally excited quasiparticles
tunnel between the layers essentially as if momentum were conserved, i.e., 
$\Lambda$ is a small scale in this regime.
In this limit we may
approximate ${\cal T}^2_{\bk-\bp}$ by a delta function and 
Eq.~(\ref{eq:lambdac}) becomes the same as the result Eq.~(\ref{eq:standardab})
for $\lambda_{ab}$, implying {\em linear} temperature dependence, 
$\delta\lambda_c(T)\sim T$.

(ii) For $\vd\Lambda\ll T \ll v_F\Lambda$ the form of ${\cal T}^2_{\bk-\bp}$ 
suggests that 
the dominant tunneling is characterized by  $\vd k_2$ being essentially conserved but 
$v_F k_1$ being essentially unrestricted. This is exactly the situation 
envisioned in Ref.\ [\onlinecite{REF:Radtke}], leading to the 
$T^2$ behavior, but in our case it emerges as a crossover phenomenon.

(iii) For  $T\ll\vd\Lambda \ll v_F\Lambda$, extending this simple argument
(so that neither component is conserved)
yields a vanishing result for $\delta \lambda_c(T)$ due to the sign-change
in the d-wave pair potential. Using a more 
careful Sommerfeld-like expansion detailed in Appendix~\ref{a2}, we 
find that $\delta \lambda_c(T) \propto T^3$ in this regime.

Summarizing, we have 
\begin{eqnarray}
\delta \lambda_c(T) \sim
\begin{cases}
T^3
       \,\,\, {\rm for} \,\,\, T\ll \vd\Lambda \ll \vf\Lambda;\\
T^2
       \,\,\,  {\rm for}  \,\,\, \vd\Lambda\ll T\ll \vf \Lambda;\\
T
     \,\,\,\,\,    {\rm for}   \,\,\,\vd\Lambda \ll\vf\Lambda\ll T.\\
\end{cases}
\label{T-dep}
\end{eqnarray}
The c-axis penetration depth data of Ref.~[\onlinecite{REF:Hosseini,prl}]
 exhibit power-law
behavior consistent with $T^3$ crossing over to $T^2$. The first two regimes
of Eq.~(\ref{T-dep}) indicate that, by a suitable choice
of the parameter $\Lambda$, it may be possible to fit this data with the present model.
Before attempting such a fit, we turn to the $\Ec$ dependence of $\lambda^{-2}_c(0)$.

 Next, we carry out a similar analysis
for $\lambda^{-2}_c(0)$, which has the explicit form of Eq.~(\ref{eq:lambdac}) but
with each $\tanh $ function replaced by unity. 
 In analyzing $\delta \lambda_c(T)$, we made use of the interplay between
the way in which integrals were cut off by $T$ and the way momenta were
effectively conserved by ${\cal T}^2_{\bk-\bp}$.  
For $\lambda^{-2}_c(0)$, the momentum integrals are cut off by $\Ec$ instead of 
$T$ (via the 
factor $Z_{\bf k}Z_{\bf p}$), but otherwise the previous analysis remains 
largely valid.   
In particular, the intermediate ($\vd\Lambda\ll \Ec\ll \vf \Lambda$) and high 
($\vd\Lambda \ll\vf\Lambda\ll \Ec$) $\Ec$ regimes are analogous to their counterparts 
in Eq.~(\ref{T-dep}).  At low $\Ec$, the naive analysis fails (as it did for 
$\delta \lambda_c(T)$ at low $T$); a more careful analysis (described in Appendix~\ref{a2}) 
shows that $\lambda_c^{-2}(0)\propto  \Ec^5$ for $\Ec \to 0$.   
We thus have  
\begin{eqnarray}
\lambda_c^{-2}(0) \sim
\begin{cases}
\Ec^5
       \,\,\, {\rm for} \,\,\, \Ec\ll \vd\Lambda \ll \vf\Lambda;\\
\Ec^2
       \,\,\,  {\rm for}  \,\,\, \vd\Lambda\ll \Ec\ll \vf \Lambda;\\
\Ec
     \,\,\,\,\,    {\rm for}   \,\,\,\vd\Lambda \ll\vf\Lambda\ll \Ec.\\
\end{cases}
\label{x-dep}
\end{eqnarray}

Our results (\ref{T-dep}) and (\ref{x-dep}) indicate that the incoherent 
tunneling model with effective charge renormalization (\ref{eq:z1})
qualitatively reproduces the trends in the $c$-axis penetration depth
summarized in Eq.\ (\ref{eq:slopec}). The near-quadratic behavior in both
temperature and doping observed in experiment
\cite{REF:Hosseini,prl} emerges
in our model as a crossover phenomenon involving the energy scales 
$\vd\Lambda \ll v_F\Lambda$. We shall see below that the full numerical
integration of Eq.\ (\ref{eq:lambdac}) indeed reproduces the 
scaling behavior indicated in Eqs.\ (\ref{T-dep}), (\ref{x-dep}) and 
furthermore gives excellent {\em quantitative} agreement with the 
experimental data.
From our analysis above it should also be clear that if $Z_\bk$ fell
to a small value proportional to $\sqrt{x}$ outside the nodal patch (instead
of vanishing there) then $\lambda_c^{-2}(0)$ would pick up a contribution 
{\em linear} in $x$ in all the regimes described in Eq.\ (\ref{x-dep}),
in disagreement with the data.

\subsection{$c$-axis penetration depth: numerical evaluation}
\label{SEC:numerical}

To facilitate numerical evaluation
it is useful to rewrite Eq.\ (\ref{eq:lambdac}) by 
converting the sums to integrations in the usual way
and switching to relative and center-of-mass variables 
${\bf q} = ({\bf k}-{\bf p})/2$ and
 ${\bf Q} = ({\bf k}+{\bf p})/2$. We have 
\begin{eqnarray}
&&\frac{1}{\lambda_c^2}
= \frac{16e^2d }{(\vf\vd)^2}
\int \frac{d^2 Q}{(2\pi\hbar)^2} \int \frac{d^2 q}{(2\pi\hbar)^2} 
Z_{{\bf Q}+{\bf q}} Z_{{\bf Q}-{\bf q}} \label{eq:caxisfin}
\\
&& \qquad\times
 {\cal T}_{2\tilde{\bf q}}^2 
\frac{q_2^2-Q_2^2}{|{\bf Q}+{\bf q}||{\bf Q}-{\bf q}|}
\frac{1}{{\bf q}\cdot {\bf Q}}
\nonumber
\\
&& \qquad\times \left[
|{\bf Q}-{\bf q}|\tanh \frac{|{\bf Q}+{\bf q}|}{2T}
-
|{\bf Q}+{\bf q}|\tanh \frac{|{\bf Q}-{\bf q}|}{2T}
\right],
\nonumber
\end{eqnarray}
where we have also linearized in the vicinity of the nodes and 
rescaled the momenta $\vf Q_1 \to Q_1$ and 
$\vd Q_2 \to Q_2$ and similarly for ${\bf q}$.  Such a rescaling effectively 
makes the argument of the matrix element 
factor anisotropic: In Eq.~(\ref{eq:caxisfin}) its argument is  
$\tilde{\bf q} \equiv (q_1/\vf, q_2/\vd)$.  

Before proceeding, we re-emphasize that incoherence in the tunneling matrix 
element is essential for 
obtaining the correct temperature dependence of $\lambda_{c}(T)$ (and, 
with our choice for $Z_{\bf k}$, the correct
doping dependence of  $\lambda_{c}(T)$).  For the case $\Lambda \to 0$, 
in which the spatially modulated 
tunneling matrix element becomes uniform on average, Eq.~(\ref{eq:caxisfin})
 yields a $T$-linear 
finite temperature correction $\lambda^{-2}_{c}(0)-\lambda^{-2}_{c}(T)$.
Thus, incoherence in the tunneling matrix element is crucial in what follows.

\begin{figure}
%
 \epsfxsize=8cm
\centerline{\epsfbox{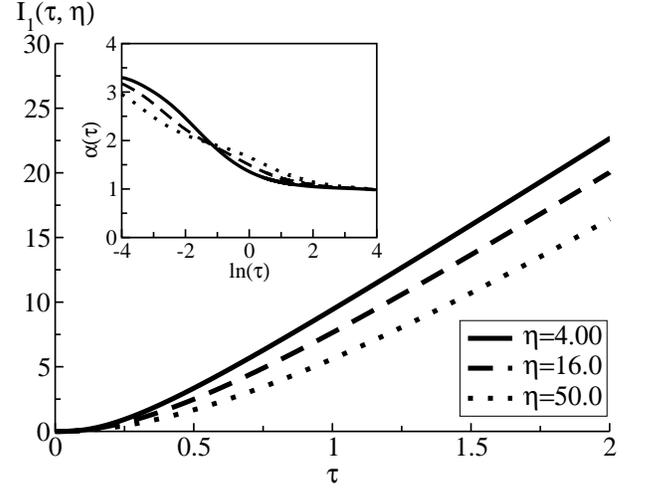}}
\vskip0.50cm \caption{Plot of $I_1(\tau,\eta)$ for $\eta = 4,16,50$.  
To emphasize the crossover behavior 
in the power-law of  $I_1(\tau,\eta)$, the inset plots the associated 
logarithmic derivative $\alpha(\tau)=d\ln{I_1}/d\ln{\tau}$.  
}
\label{fig2}
\end{figure}

It is convenient to express all quantities as dimensionful prefactors 
multiplying 
dimensionless functions of dimensionless parameters.  We express   
$\delta \lambda_c(T)$ in this manner as  
\begin{subequations}\begin{eqnarray}
\label{eq:deltalambdac}
&&\delta \lambda_c(T) =\frac{16 e^2\Lambda d} {\pi\sqrt{\vf \vd}}
\frac{Z_0^2 t_\perp^2}{h^4}
I_1(\frac{T}{\sqrt{\vf\vd}\Lambda}, \eta),
\\
&&I_1(\tau, \eta) \equiv  4\tau^3 \int d^2 q 
{\rm e}^{-4\tau^2(q_1^2/\eta - q_2^2\eta)} \Omega({\bf q}),
\label{eq:i}
\\
&& \Omega({\bf q}) \equiv  \frac{1}{4}\int d^2 Q  
\frac{q_2^2-Q_2^2}{|{\bf Q}+{\bf q}||{\bf Q}-{\bf q}|}
\frac{1}{{\bf q}\cdot {\bf Q}}
\nonumber \\
&&\qquad
\times\left[
|{\bf Q}-{\bf q}|(1- \tanh \frac{|{\bf Q}+{\bf q}|}{2})
- \right.
\nonumber \\
&&\qquad\qquad
\left. -
|{\bf Q}+{\bf q}|(1-\tanh \frac{|{\bf Q}-{\bf q}|}{2})
\right].
\end{eqnarray}\end{subequations}
For simplicity in the above integrals we have set $Z_\bk$=$Z_0$, 
an approximation valid as long as $T\ll \Ec$. The integrals remain 
convergent due to the thermal Fermi factors. 

In the main part of Fig.~\ref{fig2} we display 
$I_1(\tau, \eta)$ for $\eta = 4,16,50$; other values of $\eta$ display 
similar  behavior.  To focus on the power-law behavior of $I(\tau, \eta)$, 
in the inset 
of Fig.~\ref{fig2} we plot the logarithmic derivative 
$\alpha(\tau)=d\ln{I_1}/d\ln{\tau}$
as a function of $\ln{\tau}$, again for $\eta = 4,16,50$.  The virtue of 
such a plot is that constant behavior of the logarithmic derivative at a 
particular value reflects power-law
behavior of $I(\tau, \eta)$ with that value as the exponent.  By examining 
this plot, there is clear evidence for  $\tau$-linear behavior for 
$\tau \gg \sqrt{\eta}$ (in fact, the exponent approaches unity for much lower 
values of $\tau$) and  $\tau^3$ behavior for $\tau \ll 1/\sqrt{\eta}$.  
At intermediate $\tau$ the exponent is close to 2 although the expected 
plateau starts developing only for large anisotropy $\eta$. We may thus
conclude that $\delta\lambda_c(T)$ indeed exhibits behavior expected from 
the scaling analysis presented above.

\begin{figure}
 \vskip0.50cm
 \epsfxsize=8cm
\centerline{\epsfbox{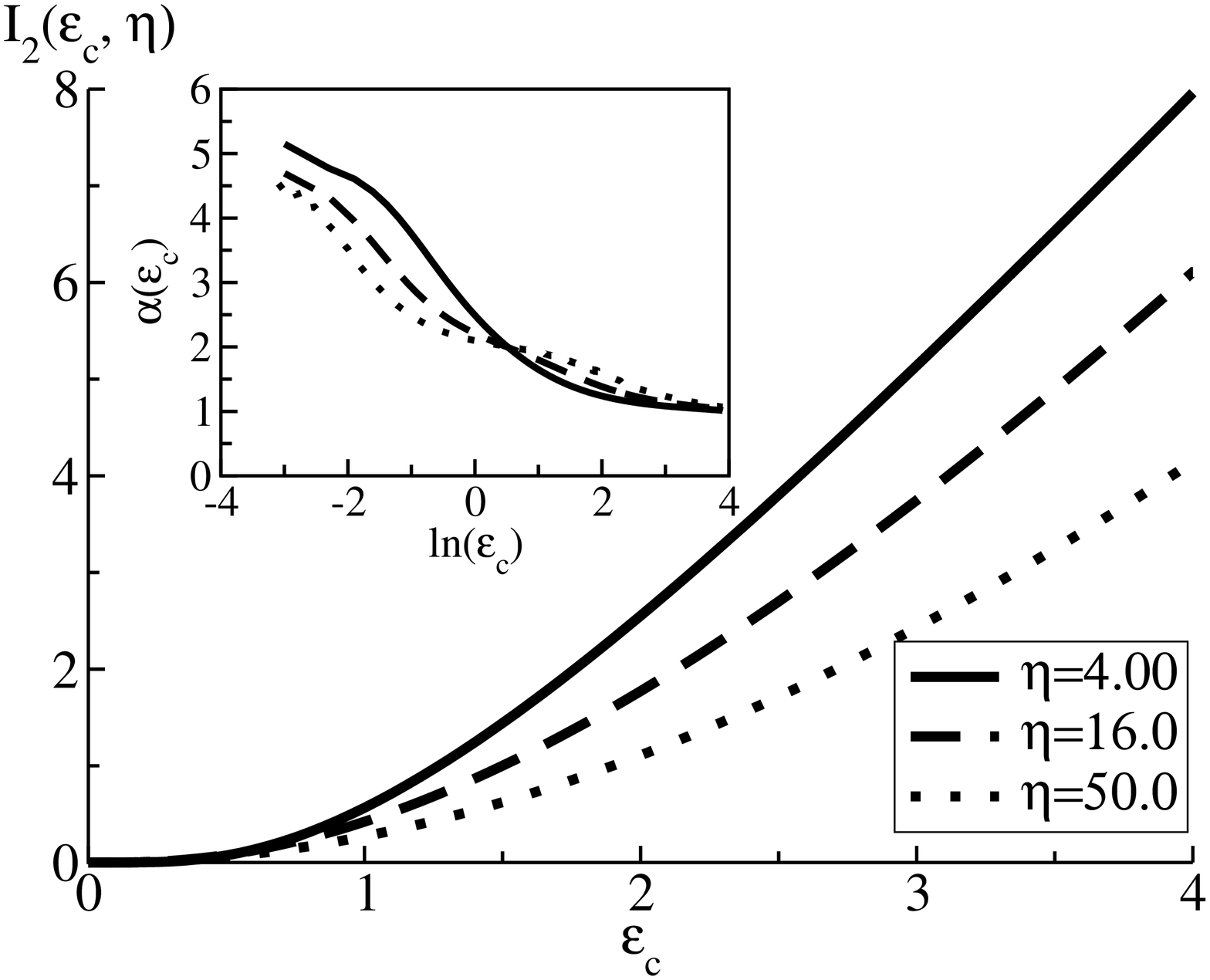}}
\vskip0.50cm \caption{Plot of $I_2(\ec,\eta)$ for 
$\eta = 4,16,50$.  To emphasize the crossover behavior 
in the power-law of $I_2(\ec,\eta)$, the inset plots the 
associated logarithmic derivative 
$\alpha(\ec)=d\ln{I_2}/d\ln{\ec}$.  
}
\label{fig3}
\end{figure}
We now analyze the doping dependence.
It is convenient to express $\lambda^{-2}_c(0)$ in a form reminiscent of 
Eq.~(\ref{eq:deltalambdac}); i.e., as a product of a dimensionful prefactor 
multiplying a dimensionless function of suitably chosen dimensionless 
quantities.  We thus have 
\begin{subequations}\begin{eqnarray}
\label{eq:lambdacn}
&&\frac{1}{\lambda_c^2(0)}=
\frac{16 e^2\Lambda d} {\pi\sqrt{\vf \vd}}
\frac{Z_0^2t_\perp^2}{h^4}
I_2(\frac{\Ec}{\sqrt{\vf \vd} \Lambda},\eta),
\\
&&I_2(\ec,\eta)=
\ec^3 
 \int_{1} d^2k \int_{1} d^2p\, \frac{k_2 p_2}{kp}\frac{1}{k+p}
\nonumber \\
&&\qquad\qquad \times{\rm e}^{-\ec^2[(k_1-p_1)^2/\eta - 
(k_2-p_2)^2\eta]},\label{eq:i2}
\end{eqnarray}\end{subequations}
where the subscript ${1}$ on the integrations in Eq.~(\ref{eq:i2}) indicate 
that the integration range is the unit disk,
corresponding to the discontinuous jump in our choice for $Z_{\bf k}$.  In 
practice, for numerical convenience we shall replace this hard 
cutoff with a Gaussian soft cutoff when performing numerical integrals.  
This corresponds to the choice   $Z_{\bf k} =Z_0 \exp(-E_{\bf k}^2/\Ec^2)$
which leads to the same qualitative $ab$-plane behavior as discussed in 
Sec.~\ref{SEC:ab}.  

To ascertain whether the three regimes indicated in Eq.\ (\ref{x-dep}) 
are indeed realized, in the main part of Fig.~\ref{fig3} we 
display a numerical plot of $I_2(\ec,\eta)$ for
$\eta = 4,16,50$.  The inset of Fig.~\ref{fig3} we plot the logarithmic 
derivative of this quantity.  
For large $\ec$, $I_2(\ec,\eta)$ clearly exhibits 
linear behavior.  For 
intermediate  $\ec$, there is clearly a regime of quadratic 
behavior that is more 
pronounced for the cases $\eta = 16$ and $\eta = 50$ and the exponent is
seen to approach 5 for low $\ec$. 
The essential feature of  $I_2(\ec,\eta)$ is that it 
exhibits power-law behavior with
an exponent near $2$ for a considerable range of $\ec$. The 
hope  is that this quadratic dependence
can, with a suitable choice of parameters, fit the known quadratic doping 
dependence of the $c$-axis penetration 
depth Eq.~(\ref{eq:slopec}).   In the next section, we perform a 
detailed fit to $c$-axis data.\cite{REF:Hosseini,prl}

%
\section{$c$-axis penetration depth: data fits}
\label{SEC:fits}
In the present section, we attempt to fit the c-axis penetration depth data of 
Ref.~[\onlinecite{REF:Hosseini,prl}] using  the theoretical formulas
obtained in the preceding section.  
Thus, we assume that the experimental
$\lambda^{-2}_c(T) \approx  \lambda^{-2}_c(0) - \delta \lambda_c(T) $ with   
$\lambda^{-2}_c(0)$
and $\delta \lambda_c(T) $ being given by Eq.~(\ref{eq:lambdacn}) and
Eq.~(\ref{eq:deltalambdac}), respectively.  Within this approximation to 
Eq.~(\ref{eq:caxisfin}), $\delta \lambda_c(T) $ does not depend on 
$E_{\rm c}$, in agreement with the 
experimental observation that the finite-$T$ correction is universal 
(i.e.~doping-independent) 
at low $T$.  Thus, we
begin by first fitting $\delta \lambda_c(T) $ to the experimental 
finite-$T$ correction
$\lambda^{-2}_c(0) - \lambda^{-2}_c(T)$.  

\begin{figure}
%
 \epsfxsize=\columnwidth
\centerline{\epsfbox{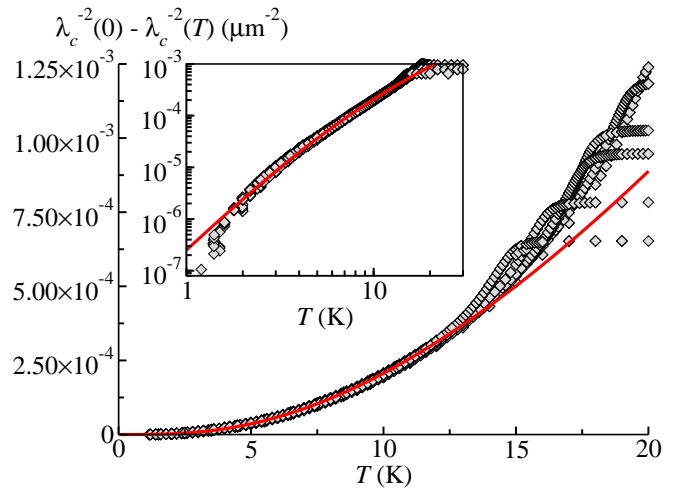}}
\caption{Plot of fits to experimental data of 
Ref.~[\onlinecite{REF:Hosseini,prl}].  The diamonds are 
$\lambda^{-2}_c(0) -\lambda^{-2}_c(T)$ for various dopings having 
experimental $\tc$ values 
(top to bottom) $20.2,19.5,18.2,17.8,16.4,15.1$K.  The solid curve is our 
best fit using the parameters $\hbar\Lambda^{-1} = 120$\AA\ and 
$t_\perp = 26$meV.  The inset is the same plot on a logarithmic scale, showing the changing
power law of the experimental data and of our theoretical curve.}
\label{fig4}
\end{figure}

The relevant fitting parameters are as follows:
The parameters $t_\perp$ and $\Lambda$ characterize the
way in which tunneling occurs in the $c$-axis direction; since the 
measurements of Ref.~[\onlinecite{REF:Hosseini,prl}] were done on 
one crystal, we shall take these to be {\it global\/}
fitting parameters. 
The parameter $Z_0$ may be taken to be unity, as it only enters in Eq.~(\ref{eq:lambdacn}) and
Eq.~(\ref{eq:deltalambdac}) in the product $Z_0 t_\perp$.
We take $\vf = 1.8$ eV \AA \cite{REF:Ioffe} and 
$d = 5.85$\AA, although it is not known 
if  $\vf$ changes for strongly underdoped samples.  As discussed above, 
the way in which the power-law behavior of Eq.~(\ref{eq:deltalambdac})  
changes as quasiparticles are excited from the condensate depends on 
the magnitude of the anisotropy ratio $\eta = \vf/\vd$; however, in 
practice we have not found the quality of the
fits to be strongly dependent on this value.  Thus, although the best 
fits (i.e.~those minimizing the variance) 
were found with the value $\eta = 6.8$, one cannot claim to extract a value for $\eta$ using this
 technique.  The converse 
of this is that although it is expected that $\eta$ will vary somewhat for such underdoped samples
 (in a way that is not understood
at present), the weak dependence of the quality of our fits on $\eta$ validates our neglecting this
 doping dependence. 

In Fig.~\ref{fig4} we show our best fits to the low-temperature values of 
$\lambda^{-2}_c(0) - \lambda^{-2}_c(T)$ in the experiments
of Ref.~[\onlinecite{REF:Hosseini,prl}].  
The diamonds represent data curves for 
a particular doping value.  Each doping value is characterized by a 
particular $\tc$ which we take to be proportional to $x$.  
For clarity we have only displayed the highest doping values
(i.e., $\tc = 20.2,19.5,18.2,17.8,16.4,15.1$K); the fits are equally good for 
lower doping values.  
The solid line is our best fit with the parameters 
$\hbar/\Lambda = 120$\AA ~and $t_\perp = 26$meV.  The fits work well at 
low $T$ (despite the fact that the data is {\it not\/} a simple power law) 
but begins to deviate at high temperatures.  
(The roughly horizontal behavior occurs above $\tc$).
We ascribe this
discrepancy to fluctuation effects near $\tc$ in a given sample, as well as 
the fact that we have neglected the effect of $Z_{\bf k}$ on
the finite temperature correction $\delta \lambda_c(T) $ in 
Eq.~(\ref{eq:deltalambdac}). This restricts the validity of our calculations
to low temperature.  As we saw in Sec.~\ref{SEC:ab} for $\lambda_{ab}$, 
the deviations due to this effect are expected to become more 
pronounced at higher temperatures (see Fig.~\ref{fig1}).

\begin{figure}
 \epsfxsize=7cm
\centerline{\epsfbox{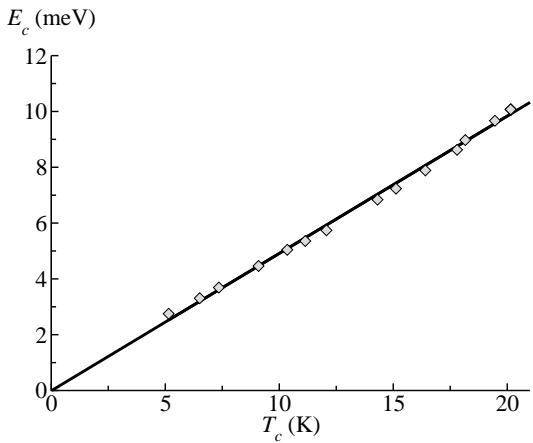}}
\caption{Plot of extracted values of the charge-renormalization parameter 
$\Ec$ (diamonds) as a function of the experimental 
$\tc$, showing a linear behavior as a function of doping level.  
The solid curve is a linear fit to these values, and has the 
form $\Ec = 0.49 \tc/{\rm K} + 0.01$(meV).
}
\label{fig5}
\end{figure}

Having fit the temperature-dependent correction, the {\it only\/} remaining 
parameters are the values of $E_{\rm c}$ corresponding to a 
particular doping.  We extract these using Eq.~(\ref{eq:lambdacn}) for 
$\lambda^{-2}_c(0)$.  
  In Sec.~\ref{SEC:ab} we noted that to account for the $ab$-plane 
phenomenology, we must have $\Ec \propto x$ (and therefore $\Ec 
\propto \tc$).  To verify that this indeed holds, in Fig.~\ref{fig5} we 
plot (diamonds) the extracted best-fit values of 
$\Ec$ for a given experimental $\tc$ from the Hosseini {\em et al.}
 results.  The solid curve  is a linear fit to these values, 
with the form  $\Ec = 0.49 \tc/{\rm K} + 0.01$(meV).  This agrees 
very well with the rough estimate of Sec.~\ref{SEC:ab}
where we found $\Ec = 0.55T_c$ meV/K  from the $ab$ plane data.
 This linear
``Uemura''~\cite{REF:Uemura} relation is an important constraint 
on this theory and depicts the destruction of the Fermi surface as the 
Mott insulating phase is approached at low doping values.

\begin{figure}
%
 \epsfxsize=8cm
\centerline{\epsfbox{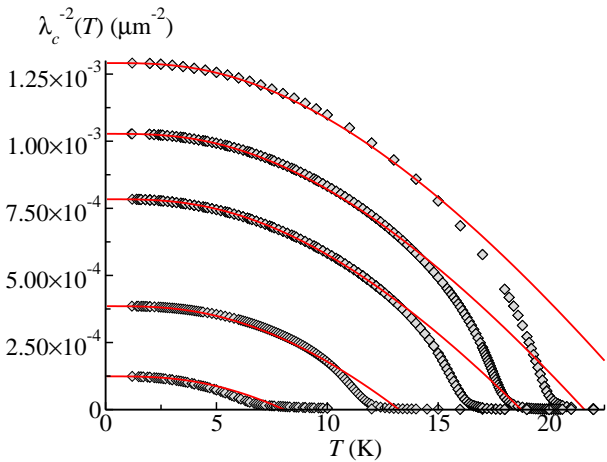}}
\caption{Fit to data of Ref.~[\onlinecite{REF:Hosseini,prl}] (diamonds)
using parameters extracted in text.  The $\tc$
values are $\tc = 20.2,18.2,16.4,12.1,7.4$K, (top to bottom) representing 
decreasing effective doping. The parameters used are  
$\hbar\Lambda^{-1} = 120$\AA\,  $t_\perp = 26$meV, $\eta=v_F/v_\Delta=6.8$
and $E_c=0.49T_c$meV/K.
}
\label{fig8}
\end{figure}

Finally, to illustrate the overall agreement of our model with the data,
in Fig.~\ref{fig8} we plot the data of Hosseini {\em et al.}
\cite{REF:Hosseini,prl} for several representative doping 
values along with our curve fits. The agreement is strikingly good
at low temperatures for all doping levels. We emphasize that all data sets
are fit with a single set of parameters (listed in Fig.\ \ref{fig8}); the 
only parameter that varies is the cutoff energy according to
$E_c=0.49T_c$meV/K with $T_c$ being the actual measured critical temperature. 

\section{Concluding Remarks}
\label{SEC:concl}

In an effort to understand $ab$-plane and  $c$-axis superfluid stiffness
in underdoped cuprates we have 
constructed a model incorporating incoherent tunneling between the 
CuO$_2$ planes along with 
a strongly anisotropic charge renormalization factor $Z_{\bf k}$.  
Incoherent tunneling provides a mechanism for obtaining a non-linear 
temperature dependence of $\lambda_c^{-2}$ while the charge renormalization
factor is introduced as means to model the fundamental departure from the 
BCS theory in the underdoped regime where only doped holes contribute
to the superfluid.

At this stage our choice for $Z_{\bf k}$ is purely phenomenological
and  expands upon the suggestions of Refs.~[\onlinecite{REF:Ioffe}]
and~[\onlinecite{REF:Millis98}]. It is motivated by the 
observation~\cite{REF:Lee} that although the $ab$-plane  and $c$-axis 
penetration depths are strongly doping dependent, the temperature-dependent 
corrections
to these doping-dependent values are nearly doping-independent.  At the 
coarsest level, this implies that quasiparticle effective charge 
$Z_{\bf k} \simeq 1$ for states near
the nodes  of the $d$-wave order parameter but is strongly reduced away 
from the nodes in a doping-dependent 
fashion. This doping dependence is governed by a cutoff energy $\Ec\sim x$.
By incorporating these two features into a model of the $ab$-plane and 
$c$-axis penetration 
depths, we were able to explain the low-temperature 
data\cite{REF:Hosseini,prl} with a striking accuracy using
a single set of input parameters and values 
of $\Ec$ which vary remarkably linearly with doping level, as expected 
on the basis of the $ab$-plane phenomenology.  

Taken together the above results lead to the notion of a ``nodal 
protectorate'' in which coherent BCS quasiparticles persist even as
the system approaches the Mott-Hubbard insulating state near the half 
filling. This nodal protectorate is schematically depicted in 
Fig.~\ref{fig6} which in addition illustrates how the protected regions shrink
to essentially nothing as $x\to 0$. The existence of this nodal protectorate is
in addition supported by heat conduction \cite{REF:Sutherland}, scanning 
tunneling spectroscopy \cite{davis1} and photoemission experiments 
\cite{norman1}.

The existence of the nodal protectorate imposes a number of stringent constraints on 
any microscopic theory describing the the underdoped regime. In particular 
any such theory must explain what protects the low-energy nodal excitations 
from the strong interactions that otherwise drive the electrons in the
remainder of the Brillouin zone inert to applied electromagnetic fields.
In addition one would like to understand what is the significance of the
energy scale $E_c\sim x$ implied by our results. It is well known that 
RVB-type theories\cite{rvb} naturally explain the $x$-linear dependence
of $\rho_s^{ab}(0)$ but predict a coefficient of the $T$-linear term
to go as $x^2$, in strong disagreement with the experimental
data. It has been claimed that the SU(2) version of the theory rectifies
this problem\cite{REF:Lee} but the physics of this is somewhat opaque
and the status of this result remains unclear.\cite{remark2} In addition
it is not obvious that one would obtain the required $c$-axis behavior
from this model. Approaches which seek to describe the Mott physics
via Gutzwiller projection techniques applied to the BCS wavefunction
\cite{projected,laughlin} 
also obtain the correct $\rho_s^{ab}(0)\sim x$ behavior. It 
is more difficult to address the finite-$T$ properties within these models
and the behavior of $\delta\lambda_{ab}(T)$ is not known at present. However, 
these being real space techniques, it is not easy to envision a mechanism 
that would protect the nodal regions from the strong correlations imposed
by the projection and one would naively expect 
that the result will suffer from the same problem as the RVB theories. 

It has been argued that the $x$ and $T$ dependence 
of  $\rho_s^{ab}$ can be explained within the Fermi liquid theory for 
superconductors with anisotropic Fermi surface\cite{walker1,arun1} assuming
a particular form of the interaction $f_{\bk\bk'}^{\sigma\sigma'}$. 
We have checked that
a most straightforward extension of this interaction does not reproduce 
the $c$-axis phenomenology; one would have to impose another set of 
constraints for the interlayer interactions to get the observed behavior 
which makes this route somewhat implausible in our view. One 
could imagine a competing order\cite{vojta1,DDW}
 in the particle-hole channel gapping out the Fermi
surface away from the nodes, thus reducing the number of electrons 
participating in the superfluid in a manner consistent with the observed
phenomenology.\cite{DDW1,kim1} 
The advantage of this scenario is that nodal quasiparticles
would automatically remain protected and the $c$-axis phenomenology would 
presumably also follow. On the other hand such competing orders necessarily 
break various space-time symmetries of the underlying system and such 
symmetry breaking should be easily observable if the competing 
order was strong enough to modify the superfluid response in accord 
with experiment. On balance we feel that the available evidence does not
support this scenario.\cite{dhlee1}

In the phase fluctuation scenarios
\cite{emery1,balents1,ft1,REF:Uemura} the low superfluid
density  $\rho_s(0)\sim x$ enters as a phenomenological input. An appealing 
feature of these models is that in the superconducting state the phase 
fluctuations are gapped
(with the gap of the order of $\rho_s\sim E_c$) and thus do not affect the 
low energy nodal quasiparticles which remain sharp even as $x\to 0$. On the
other hand the phase fluctuation models require some microscopic theory to 
describe the Mott physics that is ultimately responsible for the assumed low 
superfluid density. 

It would thus appear that none of the microscopic theories that exist 
at the present time satisfies all of the constraints implied by the 
phenomenological model
advocated in this paper. We may conclude that the physics of a $d$-wave 
superconductor on the brink of becoming a Mott-Hubbard insulator remains an 
intellectual challenge awaiting future solution.

\smallskip
\noindent
{\it Acknowledgments\/} ---  
We wish to thank D.A. Bonn, D.M. Broun, W.N. Hardy, A. Hosseini and R. Liang
for numerous valuable discussions and for providing their experimental data.  
In addition we are indebted to P.A. Lee, A.J. Millis, C. Nayak, X.-G. Wen 
and Z. Tesanovic
for discussions and correspondence. We thank I. Olabarrieta for assistance 
with numerical computations which
were performed in part on the vn.physics.ubc.ca cluster, funded
by the Canadian Foundation for Innovation (CFI) and the BC Knowledge
Development Fund.
Part of this work was carried out at the Aspen Center for Physics.
This work was supported by NSERC, CIAR and the A.P. Sloan Foundation.


\appendix
\section{Calculation of $\lambda_{ab}$ }
\label{a1}
 In the present section, we compute the in-plane penetration depth for a BCS $d$-wave superconductor 
described by the Hamiltonian $H$ in Eq.~(\ref{eq:inplaneham})
 with $H_{\rm int}=0 $ for now.  This is related to the imaginary part of the in-plane 
conductivity $\sigma_2(\omega)$ via~\cite{REF:Tinkham}
\begin{equation}
\frac{1}{\lambda_{ab}^2}=\lim_{\omega\to 0}{\omega\sigma_2(\omega)}.
\label{lam}
\end{equation}
The first step is to identify how the system couples to 
an applied electric field $E_x$ in the $x$ direction.  
For the case of $H$ having only nearest-neighbor hopping, 
the electromagnetic current operator has the form 
\begin{equation}
\label{eq:j2}
j_x({\bf r}) = ie a \sum_{\sigma} (t c_{{\bf r},\sigma}^{\dagger}
c_{{\bf r}+a\hat{x},\sigma}^{\phdag} - {\rm h.c.} ),
\end{equation}
with $a$ being the lattice spacing.  The coupling to 
the electromagnetic vector potential 
 $A_x$ (related 
to $E_x$ via
$A_x = -\frac{ic}{\omega}E_x {\rm e}^{-i\omega t} $) enters by making the 
Peierls replacement $t \to t\exp (iea A_x/c)$ in Eq.~(\ref{eq:j2}).
Expanding to to leading order in $A$, we have   
$j_x = j^{\rm p}_x + j^{\rm d}_x$ with the paramagnetic and diamagnetic 
currents being given, respectively, by 
\begin{eqnarray}
j^{\rm p}_x &=&
 iea t \sum_{\sigma} ( c_{{\bf r},\sigma}^{\dagger}
c_{{\bf r}+a\hat{x},\sigma}^{\phdag} - {\rm h.c.} ),
\\
j^{\rm d}_x &=&
 -\frac{e^2 a^2 t}{c}  \sum_{\sigma} ( c_{{\bf r},\sigma}^{\dagger}
c_{{\bf r}+a\hat{x},\sigma}^{\phdag} +{\rm h.c.} )A_x .
\label{eq:dia}
\end{eqnarray}
Our task is to find the conductivity $\sigma_{\rm L}(\omega)$ for a single layer,
  which is defined by
$\langle j_x(q,\omega)  \rangle = \sigma(q,\omega) E_x(q,\omega)$. 
The conductivity of the layered 
cuprate system will then be $\sigma(\omega)= \frac{n}{d}\sigma_{\rm L}(\omega)$ with $n$ being the number of CuO$_2$
planes per layer and $d$ being the interlayer spacing.   The angle brackets
represent the equilibrium average with respect to $H$, and henceforth we 
shall be concerned only with 
the ${\bf q} \to {\bf 0}$  limit.  In this limit, the Fourier transformed 
paramagnetic and diamagnetic current operators may be written as 
\begin{eqnarray}
j^{\rm p}_x &=&
 e \sum_{{\bf k},\sigma} \frac{\partial\epsilon_{\bf k}}{\partial k_x}
 c_{{\bf k},\sigma}^{\dagger} c_{{\bf k},\sigma}^{\phdag} ,
\\
j^{\rm d}_x &=&
 \frac{ie^2}{\omega} \sum_{{\bf k},\sigma} \frac{\partial^2\epsilon_{\bf k}}{\partial k_x^2}
 c_{{\bf k},\sigma}^{\dagger} c_{{\bf k},\sigma}^{\phdag} E_x ,
\label{eq:dia2}
\end{eqnarray}
%
%
%
where we have displayed the expression for  general $\epsilon_{\bf k}$ 
(and will continue to do so henceforth). The frequency dependent conductivity
per layer is 
\begin{equation} 
\sigma_{\rm L}({\bf 0},\omega) = \frac{i}{\omega} [D + \Pi^{\rm R}(\omega)], 
\end{equation}
where $D$ and $\Pi$ represent the diamagnetic and paramagnetic parts of the 
response. 
As usual, the diamagnetic current is already linear in the electric field so 
that the associated
contribution to the conductivity can be directly read off by taking the 
expectation value of Eq.~(\ref{eq:dia2}),  
\begin{eqnarray}
&&D=  e^2 \sum_{{\bf k},\sigma} \frac{\partial^2\epsilon_{\bf k}}{\partial 
k_x^2}
 \langle c_{{\bf k},\sigma}^{\dagger} c_{{\bf k},\sigma}^{\phdag}\rangle,
\\
&&\quad
= 2e^2  T\sum_{i\omega}\sum_{\bf k} \frac{\partial^2\epsilon_{\bf k}}
{\partial k_x^2}G({\bf k},\omega).
\end{eqnarray}
For the paramagnetic current, the usual leading-order perturbation theory 
result~\cite{REF:Mahan} naturally leads to considering the current-current 
correlator
\begin{eqnarray}
&&
\Pi(\nu) = -\int_0^{\beta} d\tau {\rm e}^{-i\nu \tau}\langle T_{\tau} 
j^{\rm p}(\tau)j^{\rm p}(0)\rangle,
\\
&&
\qquad= 2e^2  T\sum_{i\omega}
\sum_{\bf k}
 \left( \frac{\partial\epsilon_{\bf k}}{\partial k_x}\right)^2 
 \left[G({\bf k},\omega) G({\bf k},\omega- \nu)\right.
\nonumber
\\
&&\left.
\qquad\qquad +F({\bf k},\omega) F({\bf k},\omega-\nu)
\right],
\end{eqnarray}
where the in-plane Matsubara Green functions $G({\bf k},\omega)$ and  
$F({\bf k},\omega)$ are given 
by the usual expressions
\begin{eqnarray}
G({\bf k},\omega) &=& \frac{i\omega + \epsilon_{\bf k}}{(i\omega)^2-
E_{\bf k}^2},
\label{eq:g}
\\
F({\bf k},\omega) &=& -\frac{\Delta_{\bf k}}{(i\omega)^2-E_{\bf k}^2},
\label{eq:f}
\end{eqnarray}
and $E_{\bf k} \equiv \sqrt{\epsilon_{\bf k}^2+\Delta_{\bf k}^2}$.
Performing the required Matsubara sums and  
combining the preceding expressions with Eq.\ (\ref{lam}), we find 
the penetration depth 
\begin{eqnarray}
&&\frac{1}{\lambda^2_{ab}}
= \frac{e^2n}{d} \sum_{\bf k} \frac{\partial^2\epsilon_{\bf k}}{\partial k_x^2}
\left(1 - \frac{\epsilon_{\bf k}}{E_{\bf k}} \tanh\frac{1}{2}\beta E_{\bf k}
\right) \nonumber \\
&&\qquad\quad-\frac{e^2n}{2d}\beta \sum_{\bf k} \left( 
\frac{\partial\epsilon_{\bf k}}{\partial k_x}\right)^2 
{\rm sech}^2\frac{1}{2}\beta E_{\bf k},
\label{eq:second}
\end{eqnarray}
with $\beta=1/T$ the inverse temperature. This last expression agrees 
with the results for $\lambda_{ab}$ in the lattice model of a $d$-wave
superconductor obtained by other authors.\cite{scalapino1,dhlee1}

As written, the diamagnetic term in Eq.\ (\ref{eq:second}) has contributions
from the entire Brillouin zone while the paramagnetic term is dominated
by the nodal regions ($E_\bk\to 0$). For our purposes it will be 
convenient to recast the former into a form where it is also dominated 
by the nodal regions. This can be accomplished by integrating by parts in the
first term of Eq.\ (\ref{eq:second}). We obtain
\begin{eqnarray}
\frac{1}{\lambda^2_{ab}}
&=& \frac{e^2n}{d} \sum_{\bf k}\left[
\left(\frac{\partial\epsilon_{\bf k}}{\partial k_x}\right)^2
\frac{\Delta_\bk^2}{E_\bk^2}-
\frac{\partial\epsilon_{\bf k}}{\partial k_x}
\frac{\partial\Delta_{\bf k}}{\partial k_x}
\frac{\Delta_\bk\epsilon_\bk}{E_\bk^2}\right]
 \nonumber \\
&&\qquad\quad \times\left[\frac{1}{E_\bk}-\frac{\partial}{\partial E_\bk}
\right] \tanh\frac{1}{2}\beta E_\bk .
\label{eq:third}
\end{eqnarray}
This expression is mathematically equivalent to Eq.~(\ref{eq:second})
and gives $\lambda_{ab}$ as a $k$-space sum dominated by the nodal regions. In
addition, Eq.\ (\ref{eq:third}) has the desirable property of explicitly 
yielding zero superfluid stiffness in the normal limit, $\Delta_\bk=0$.
Finally we notice that the second term in the first line of Eq.\ 
(\ref{eq:third}) is generally small for a $d$-wave gap. In particular
it vanishes identically in the nodal approximation that we employ in 
Sec.~\ref{SEC:ab}.  Thus, we shall ignore this term.

\section{Asymptotic behavior of $\delta \lambda_c(T)$ and  $\lambda^{-2}_c(0)$ }
\label{a2}
In the present section we sketch the derivation of the low $T$ behavior of
$\delta \lambda_c(T)$ and the low $\Ec$ behavior of $\lambda^{-2}_c(0)$, i.e., the 
first lines of Eq.~(\ref{T-dep}) and  Eq.~(\ref{x-dep}), respectively.  
For $Z_{\bf k}$ we use the smooth form $Z_{\bf k}  = Z_0\exp(-E_{\bf k}^2/\Ec^2)$ introduced
in Sec.~\ref{SEC:numerical}.
We will omit
unimportant prefactors and, for simplicity, set $\vf =  \vd = 1$.  It will be clear from the 
derivation that relaxing this last assumption will not change the results.  We begin
with $\delta \lambda_c(T)$, which is given by
\begin{eqnarray}
\delta \lambda_c(T) &\propto& 
\int \frac{d^2 p}{(2\pi\hbar)^2} \frac{d^2 k}{(2\pi\hbar)^2} 
{\rm e}^{-({\bf k}-{\bf p})^2/\Lambda^2} \frac{k_2 p_2}{kp}
\nonumber \\
\label{eq:deltalamscale}
&& \times
\frac{p(1-\tanh(k/2T))}{p^2-k^2} .
\end{eqnarray}
Henceforth, the polar coordinate expressions of ${\bf k}$ and ${\bf p}$ are 
$(k,\theta)$ and $(p,\phi)$, respectively. 
Our strategy is to utilize the following Sommerfeld expansion:
\begin{eqnarray}
&&\int_0^{\infty} dk f(k) (1-\tanh k/2T ) \label{eq:fdef}
\\
&&\quad \approx  \int_0^{\infty} dk [f(0)+kf'(0)+\cdots] 
(1-\tanh k/2T ), \nonumber
\end{eqnarray}
which relies on the sharpness of the function $1- \tanh k/2T$ near $k = 0$ for $T\to 0$ and 
is valid provided that the function $f(k) $ is sufficiently smooth near $k=0$.  In the present case,
the function $f$ is given by (again, up to numerical prefactors)
\begin{equation}
f(k) = \int  dp\, d\phi\, d\theta 
\frac{kp^2\sin\theta \sin\phi}{p^2-k^2} {\rm e}^{-({\bf k} - {\bf p})^2/\Lambda^2}.
\label{eq:fk}
\end{equation}
The first thing to note is that, due to the factor of $k$ in Eq.~(\ref{eq:fk}) (in turn arising from the
$k$ in the measure of Eq.~(\ref{eq:deltalamscale})), $f(0) = 0$.  In addition, $f'(0) = 0$.  The easiest
way to see this is to note that, among the terms in $f'(k)$, the only one that can possibly be nonzero at $k=0$ 
is the one for which the derivative acts on the $k$ in the numerator of the fraction in 
Eq.~(\ref{eq:f}). Thus, 
\begin{equation}
f'(0) = \int  \,dp\, d\phi\, d\theta \sin\theta \sin\phi\, 
{\rm e}^{-p^2/\Lambda^2}=0,
\label{eq:fp}
\end{equation}
as the angular integrals vanish by symmetry.  Clearly, we must 
consider $f''(0)$.  A direct evaluation of $f''(0)$ does indeed yield a finite result, so that 
the leading contribution comes from the third term in Eq.~(\ref{eq:fdef}).  Evaluating the integral
over $k$ in this term then yields $\delta \lambda_c(T) \propto T^3$ for $T\to 0$ yielding the 
first line of Eq.~(\ref{T-dep}).

Next, we turn to the $\ec$ dependence of $\lambda^{-2}_c(0)$. 
 It is simplest to consider the scaled function $I_2(\ec,\eta)$, which
may be written as 
\begin{eqnarray}
I_2(\ec,1)&=&
 \int d^2k \int d^2p\, \frac{k_2 p_2}{kp}\frac{1}{k+p} {\rm e}^{-k^2/\ec^2}  {\rm e}^{-p^2/\ec^2} 
\nonumber \\
&&\times{\rm e}^{-({\bf k} -{\bf p})^2}. \label{eq:eps2} 
\end{eqnarray}
Equation~(\ref{eq:eps2}) is obtained from Eq.~(\ref{eq:i2}) by replacing the integrals over the unit disk 
with a smooth Gaussian cutoff and then rescaling ${\bf k} \to {\bf k}/\ec$ (and similarly for ${\bf p}$). 
We have also set $\eta = 1$ for convenience.  
We proceed in the same manner as for the case of $\delta \lambda_c(T)$ above, writing 
$I_2(\ec,1)$ as 
\begin{eqnarray}
\label{eq:i2approx}
I_2(\ec,1) &=& \int_0^{\infty} dk g(k)  {\rm e}^{-k^2/\ec^2} ,
\\
&\approx&  \int_0^{\infty} \!dk [g(0)\!+\! kg'(0) \!+\cdots] 
 {\rm e}^{-k^2/\ec^2},
\label{eq:i2approxp}
\\
 g(k) &\equiv& \int p\,dp\, d\phi\,  d\theta  \frac{k\sin \theta \sin\phi}{k+p} {\rm e}^{-p^2/\ec^2}
\nonumber \\
&&\qquad \qquad \quad\times
{\rm e}^{-({\bf k} -{\bf p})^2}. 
\end{eqnarray}
Owing to the fact that the Gaussian function is sharply peaked near $k = 0$, we have in
Eq.~(\ref{eq:i2approxp})
 approximately evaluated
Eq.~(\ref{eq:i2approx}) by Taylor expanding $g(k)$ in analogy with  Eq.~(\ref{eq:fdef}).  As in the 
previous case, $g(0) = 0$ and $g'(0) =0$, requiring the evaluation of $g''(0)$.  Thus, the calculation  
proceeds exactly as above, with one important difference: The function  $g(k)$ contains a factor 
${\rm e}^{-p^2/\ec^2}$ in the integrand.  It is straightforward to verify that 
the integration over $p$ then gives
 $g''(0) \propto \ec^2$.  Combining this with a 
factor of $\ec^3$ arising from the third term in Eq.~(\ref{eq:i2approxp}), we have that 
$I_2(\ec,1)\propto \ec^5$ for $\ec \to 0$, immediately giving the first line of 
Eq.~(\ref{x-dep}).



\begin{thebibliography}{10}
\bibitem{remark0} See however Ref. \onlinecite{tami1} which suggests
doping dependence closer to $x^2$ in moderately underdoped YBCO single 
crystals.
\end{thebibliography}
\end{document}